\documentstyle[epsfig]{H99}

\seceqnum      

\begin{document}

\title[Approaching a homogeneous galaxy distribution]{Approaching a homogeneous galaxy distribution: results from the Stromlo-APM redshift survey}
\author[S.J. Hatton]{Steve Hatton\thanks{hatton@iap.fr}\\
Department of Physics, University of Durham, Science
Laboratories, South Rd, Durham DH1 3LE.  \\
Present address: Institut d'Astrophysique de Paris, 98bis Boulevard Arago, 
75014 Paris, France.
}

\maketitle

\begin{abstract}
Recent results from a number of redshift surveys suggest 
that the Universe is well described by an inhomogeneous, 
fractal distribution on the largest scales probed.  
This distribution has been found to have 
fractal dimension, $D$, approximately equal to $2.1$,  
in contrast to a homogeneous 
distribution in which the dimension should approach 
the value $3$ as the scale is increased.  
In this paper we demonstrate that estimates of $D$, based 
on the conditional density of galaxies, are prone to bias 
from several sources.  These biases generally result in 
a smaller measured fractal dimension than the true 
dimension of the sample.  We illustrate this behaviour in 
application to the Stromlo-APM redshift survey, showing that 
this dataset in fact provides evidence for fractal dimension 
increasing with survey depth.  On the largest scale probed, 
$r \approx 60 \hmpc$, we find evidence for a distribution 
with dimension $D = 2.76 \pm 0.10$.  A comparison between 
this sample and mock Stromlo-APM catalogues taken from 
\nbody simulations (which assume a CDM cosmology) 
reveals a striking similarity in the 
behaviour of the fractal dimension.  Thus we find no 
evidence for inhomogeneity in excess of that expected from 
conventional cosmological theory.  
We consider biases affecting future large surveys and 
demonstrate, using mock SDSS catalogues, that this survey 
will be able to measure the fractal dimension on scales at which 
we expect to see full turn-over to homogeneity, in an accurate 
and unbiased way.

\end{abstract}

\begin{keywords}
cosmology: theory -- large-scale structure of Universe -- galaxies: clustering.
\end{keywords}


\section{Introduction}
\label{sec:intro}

The completion of several large redshift surveys has provided 
us with the ability to test one of the keystones of modern 
cosmological thought, the expected homogeneity of the Universe 
on large scales.  Recent debate on this subject \cite{Guzzo97} 
has centred around reports of a fractal distribution of galaxies, 
such that the measured density of the Universe does not tend 
to a well defined average as the volume measured increases.  
Some results appear to support this hypothesis 
(\citeNP{CPS88,LEDA96}; \citeNP{S-L98}, hereafter SLM); 
others do not \cite{ESPV98}.

In a fractal Universe, the density measured 
by any observer decreases with the size of the sample 
considered.  This trend produces a bias in conventional, 
correlation function analysis of galaxy clustering, 
and is the motivation for employing the statistic known 
as the conditional density.  This quantity is defined in 
section~\ref{sec:frac}, and its relationship to the 
correlation function is explained.  In section~\ref{sec:bias}, 
we show that there exist a number of effects which can bias 
this statistic.  Systematic effects can create the 
illusion of a density which decreases with scale, even when 
the underlying distribution is homogeneous.  We examine 
the Stromlo-APM redshift survey \cite{Loved92,Loved96b} in the 
light of these biases in section~\ref{sec:stromlo}.  Previous 
work (SLM) has found that the fractal distribution in 
this catalogue is a constant function of scale, with 
no turnover to homogeneity suggested.  We show that 
a careful treatment of the biases enables us to reliably 
probe slightly deeper scales than those of SLM, and that 
the fractal dimension clearly increases with scale.  
A comparison is made between the fractal dimension 
measured from this dataset, and that obtained by analysis 
of mock Stromlo-APM catalogues taken from \nbody simulations.  
These simulations incorporate a more ``traditional'' view of 
galaxy clustering, one in which the amplitude of large-scale 
anisotropy is given by a CDM model for the galaxy power spectrum.  
In section~5 we consider the application of the same techniques 
to larger surveys, in particular the PSCz \cite{PSCz} and the 
Sloan Digital Sky Survey \cite{GW95}.  
We conclude in section~\ref{sec:conc}.  


\section{Fractal correlations}
\label{sec:frac}  
If clustering in the Universe is described by a fractal, it 
is not valid to use the conventional definition of the correlation 
function, 
\[
\xi(r) = \frac{\langle n({\bf r}) n({\bf r}+{\bf x}) \rangle}{\langle n \rangle^2} - 1 ,
\]
to examine this clustering on scales where the correlation 
function is small.  This is because, on these scales, the 
shape of this function is affected by the value of $\< n \>$, 
which has to be estimated from the sample.  
For a fractal distribution, the measured density will be 
a decreasing function of the distance from the observer, so 
there is no well defined $\< n \>$, and the shape of the correlation 
function will depend on the size of the sample.  

The conditional density, $\Gamma(r)$, is defined by
\[
\Gamma(r) = \frac{\langle n({\bf r}) n({\bf r}+{\bf x}) \rangle}{\langle n \rangle} 
\]
\cite{CP92}.  Whilst the normalization of this function depends on the 
mean density of galaxies in the sample, its shape does not.  
The correlation function is simply related to $\Gamma(r)$: 
\[
\xi(r) = \frac{\Gamma(r)}{\langle n \rangle} - 1.
\]
It is clear that in the regime of strong clustering, \ie 
$\xi(r) \gg 1$, we expect to see $\Gamma(r) \approx  \< n \> \xi(r)$.  

A fractal distribution with dimension $D$ has the property that the 
spherically averaged density around an observer obeys the scaling law: 
\[
N(r) \propto r^D.  
\]
So the conditional density is given by
\[
\Gamma(r) = A r^{D-3}
\]
where $A$ is a constant for a particular distribution.   
On small scales, then, we expect to see clustering 
with a fractal dimension $D = 3 - \gamma$, where $\gamma$ is the 
power law slope of the correlation function, generally found to 
be $1.8$ \cite{Davpee83}.  This $D=1.2$ dimensionality is indeed 
observed on the smallest scales ($r\leq 3.5\hmpc$), with 
a turnover to a fractal dimension of $\approx 2.2$ observed for  
intermediate scales \cite{Guzzo91}.  It should be noted that 
this steep correlation function on small scales is only observed 
when the real-space correlation function is measured, either 
from the angular correlation function or by correcting 
redshift-space data is some way.  The redshift-space correlation 
function is suppressed on small scales by non-linear 
peculiar velocities.  The magnitude of this effect 
increases as we look at smaller separations, and so the 
resulting power-law slope is less steep than that in real space.   
For example, \citeN{FisherIRAS194} find a power-law slope of 
$\gamma = 1.66$ in real space, as opposed to $\gamma = 1.28$ in 
redshift space.

$\Gamma (r)$ is the density in concentric shells of radius
$r$ around a point.  In order to compare regions of space 
at different distances from the observer, it is necessary 
to take a constant density, volume limited sample from 
a redshift survey.  This involves discarding a large 
fraction of the observed galaxies, thus resulting in a 
rather noisy statistic.  For this reason, we will 
work with the integrated conditional density:
\[
\Gamma^{*}(r) = \frac{3}{4\pi r^3} \int_0^r 4\pi x^2 \Gamma(x) dx.  
\]
This function effectively represents the average density in concentric 
spheres around the observer.  This integration removes noise from 
the estimate, at the expense of introducing a smoothing and 
effectively masking any change in the shape of the underlying 
function.  However, over a range of scales for which 
the distribution is well-described by a single fractal dimension,  
this statistic will tend towards the same power-law slope, 
\[
\Gamma^{*}(r) = \frac{3 A}{D} r^{D-3}
\label{eqn:gamma_star}
\]

\subsection{The maximum scale}
\label{sec:rmax}
\begin{figure}
\centerline{\epsfxsize = 8.0 cm \epsfbox{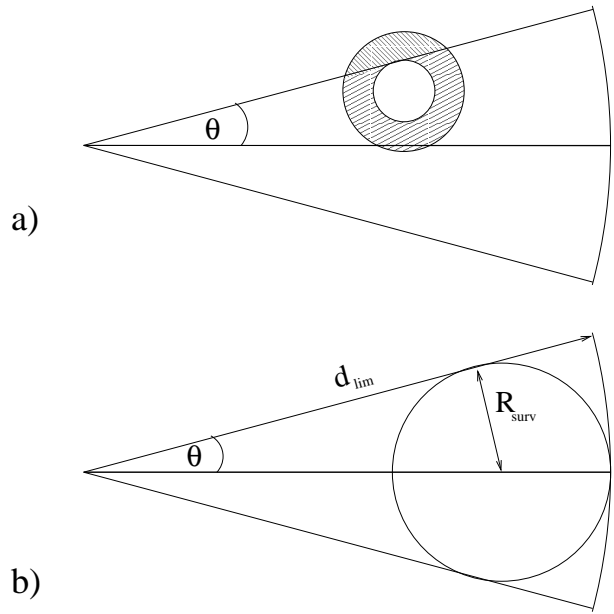}}
\caption{Schematic of a survey wedge.  The upper panel illustrates 
the state of concentric shells around a galaxy in the survey.  As 
the radius of the shell increases, eventually density is measured 
for shells that are partially external to the survey, such as the shaded 
shell above.  The lower panel demonstrates the maximum distance to 
which a given survey can probe, the radius of the largest sphere 
which can totally be contained within the survey. }
\label{fig:wedges}
\end{figure}

As we consider spherical shells around a galaxy at increasing 
radii, it is clear that we will eventually hit the edge of the 
survey.  Further increases in the shell radius will result in 
measurements of the density for shells that are underpopulated 
relative to the mean density.  This effect is displayed in 
the upper panel of Fig.~\ref{fig:wedges}.  
For the outer, shaded shell in this 
diagram, the measured density will be lower than the true 
density since it probes a region that by definition contains no 
galaxies.  In conventional correlation function analysis, this 
effect is corrected for by normalizing to the volume of the 
shell in question that {\em is} contained within the survey.  This 
is not appropriate for a fractal distribution, since the assumption 
is made that the distribution inside the survey region is the same 
as that outside.  This biases the correlation function, 
and employing it will tend to mask evidence of a fractal signal.  
The extent of this bias is debatable.  \citeN{PGM94} find that 
boundary effects alone cannot mask a true fractal distribution to 
the extent that the correlation function is as stable and homogeneous 
on large scales as has been observed.  However, the correlation 
function certainly {\em is}  biased by 
boundary effects, the only question being exactly how much.  Given 
this point, the motivation exists for adopting the most prudent 
approach, and we follow SLM in using the $\Gamma(r)$ estimator rather 
than the correlation function.

For a galaxy $i$ we define $r_{\rm max}^i$, the distance to the 
edge of the survey.  Only other galaxies with separation from $i$ 
less than this distance may be used in the pair counts that contribute 
to the estimate if the conditional density.  This requirement limits 
the range over which a given survey 
can measure $\Gamma(r)$; the maximum scale is dictated by the galaxy with 
the largest $r_{\rm max}$.  This is equivalent to the radius of the 
largest sphere that can be contained within the survey geometry, 
and depends on the opening angle of the survey, $\theta_{\rm surv}$, 
and the volume limit, $d$:
\[
R_{\rm surv} = \frac{d \, \sin \theta_{\rm surv}}{1 + \sin \theta_{\rm surv}}
\]
This is illustrated in the lower panel of Fig.~\ref{fig:wedges}.  

For a good test of homogeneity, a survey must therefore possess 
a large opening angle and reasonable depth.

\section{Biases}
\label{sec:bias}
\subsection{Small-scale cut-off}
\label{sec:poiss}
\begin{figure}
\centerline{\epsfxsize = 8.0 cm \epsfbox{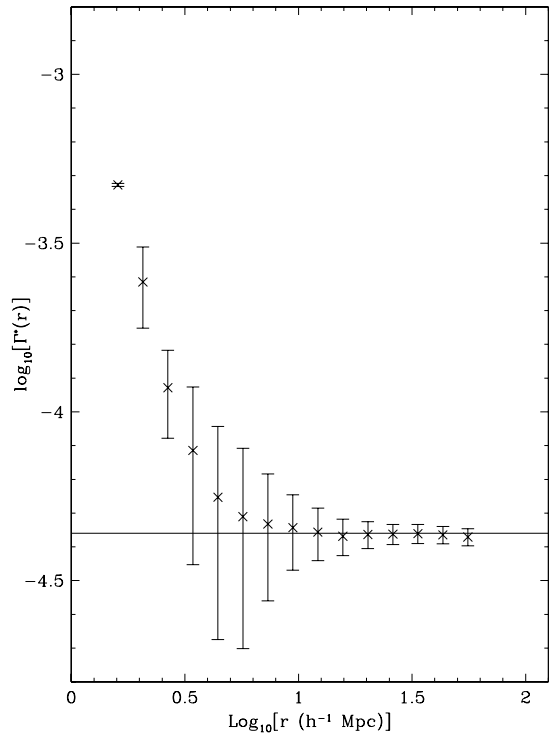}}
\caption{The behaviour of $\Gamma^{*}(r)$ for a sparse, 
$D=3$ (Poisson) sample of points.  The horizontal line 
is the large-scale asymptote. }
\label{fig:poiss}
\end{figure}

The estimator for $\Gamma^{*}(r)$ is biased on small scales 
due to the Poisson nature of the galaxy density field.  
In Fig.~\ref{fig:poiss} we demonstrate this effect.  We have 
constructed Poisson samples of galaxies and measured $\Gamma^{*}(r)$ 
down to small scales.  The points and errorbars represent averages 
and standard deviations over fifty independent realizations.  
We obtain the expected asymptotic behaviour 
on large scales for a flat, $D=3$ power law.  On smaller scales, 
however, the deviation from this homogeneous behaviour is marked.  
Thus, the signature of a homogeneous distribution can be erased 
if we attempt to fit a power law down to too small a scale, 
and it will be necessary to apply a cut-off in our fits.  

The precise scale at which the departure from fractal behaviour 
occurs is not trivial to compute, and will decrease as the 
fractal dimension of the sample decreases, since a lower fractal 
dimension results in more clustering and closer pairs of galaxies. 
However, if one is testing the existence of a turn-over to a 
homogeneity, it seems prudent to use the small-scale cut-off found 
for a homogeneous distribution, to remove any potential bias.   
We will discuss the value chosen for this cut-off in 
section~\ref{sec:cutoff}.  

\subsection{Non-constant density}
\label{sec:vl_fx}
To perform the fractal analysis, it is important to construct 
a sample with constant density.  This is achieved by selecting 
a volume limited subsample of the redshift survey in question.  
To construct a volume limited sample, we discard galaxies at  
redshifts greater than some maximum redshift, $z_{\rm lim}$.  
For each galaxy left in the survey, we define a maximum 
redshift, $z_{\rm max}$, which is the redshift that a galaxy 
with that magnitude could be placed at and still make it 
into the magnitude limit of our catalogue.  Galaxies are only 
kept if their $z_{\rm max}$ value is greater than $z_{\rm lim}$, 
\ie they could be seen if they were placed at the limiting 
redshift of the sample.   
Several effects can lead to problems with this volume limiting.  
Failing to correct for them will thus lead to a non-constant 
density over the sample, which can bias estimates of the fractal 
dimension.  These effects include:

\begin{enumerate}
\item Cosmology.  
\begin{figure}
\centerline{\epsfxsize = 8.0 cm \epsfbox{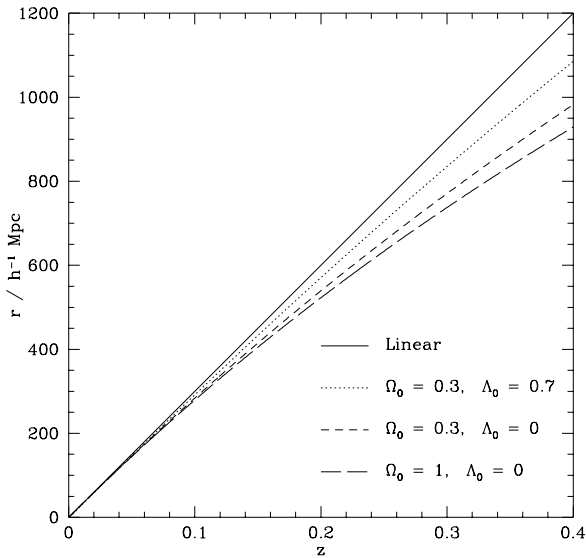}}
\caption{The linear redshift-distance compared with that for 
three cosmologies, $(\Omega_0,\Lambda_0) 
= (1,0) \, \hbox{,} (0.3,0.7) \, \hbox{and} \, (0.3,0)$ }
\label{fig:euclid}
\end{figure}
Observations of galaxy positions are made in redshift 
coordinates, but to measure the clustering in three 
dimensions we must convert to a Cartesian, comoving 
coordinate system.  This requires knowledge of the function 
$r(z)$, which depends on the cosmological parameters $\Omega_0$  
(mass density) and $\Lambda_0$ (cosmological 
constant).  In Fig.~\ref{fig:euclid} we display $r(z)$ 
for three cosmologies, and compare with the linear relation that 
is accurate for $z \ll 1$.  

The redshift-distance relation also 
affects the construction of a volume limited subsample of galaxies.  
We must define a 
maximum redshift at which each galaxy could be placed and still 
be within the catalogue magnitude limit.  This redshift is 
calculated via the relation:
\[
m-m_{\rm lim} = 5 \log_{10}[(1+z)r(z)] - 5 \log_{10}[(1+z_{\rm max})r(z_{\rm max})] 
\label{eqn:dmod}
\]

It is clear that, in general, if the wrong $r(z)$ is 
used, we will not produce a constant-density sample 
of galaxies, and the fractal dimension will be biased.

Note that, in a Universe that is truly described by a fractal 
distribution of matter, the assumption of homogeneity made in 
deriving the Friedmann equation \cite{PJE93} is no longer 
valid.  Thus, $r(z)$ calculated for any `standard' (Friedmann) 
cosmology will be inaccurate here, and new formulae for the 
metric must be derived (although \citeN{DSL98} present an 
alternative scenario whereby the mass distribution is homogeneous, 
but the galaxy distribution fractal).  
Thus, it is possible that an apparent flattening of the 
conditional density could occur in an inhomogeneous 
Universe if $r(z)$ from standard cosmology is used.  
More obviously, it can be seen that employing the wrong 
cosmological parameters in a homogeneous Universe can 
lead to a spurious fractal signature.  

\item Dust.  
Some galaxies will be thrown out of a volume limited sample 
if they are in high-extinction areas of sky, since their 
$z_{\rm max}$ values appear lower than the volume limit.  
These galaxies are in fact needed to produce a sample with 
constant density.  It is possible to correct galaxy magnitudes 
using maps of galactic extinction \cite{BH82,SFD98}.  An additional 
effect that cannot be corrected for is the suppression of observed 
galaxy number counts in the high-extinction regions of the sky.  
Galaxies have been missed by the original survey that are again 
needed if the volume limited sample is actually to be of constant 
density.  Both these effects will result in a depletion of 
galaxies at large distances.  Thus, even in a homogeneous 
distribution, the galaxy number density will be seen to fall 
on the largest scales, a spurious sign of fractal clustering.  

\item K-correction.  As galaxies are redshifted we see a different 
part of their spectrum.  Thus their apparent magnitudes depend 
on redshift and the shape of this spectrum.  The spectral shift 
generally results in an increase in the apparent magnitude of the 
galaxy.  Thus the inferred luminosities of the galaxies, if we 
fail to take this k-correction into account, are lower than their 
true, rest-frame luminosities.  In this case, a volume limited 
sample will become less dense with distance from the observer, 
and will tend to underestimate the fractal dimension.  
\citeN{ESPV98} have examined the fractal dimension of 
the ESO slice project, a redshift survey of mean depth 
$z \approx 0.1$, and find that the choice of k-correction 
can significantly bias the results for this sample.  

\item Evolution.  Similarly, the intrinsic evolution of a galaxy 
sample with look-back time results in a different class of objects 
being selected as the depth of the survey increases.  This can 
again cause variations from uniform density in a volume limited sample.  
At high enough redshift, the effect of galaxy mergers 
will ultimately increase the number density on the largest scales.  
\end{enumerate}

\subsection{Errors in $r_{\rm max}$}
As explained in section~\ref{sec:rmax}, for each galaxy we only 
include neighbours out to the distance to the edge of the 
survey.  If we overestimate the value of $r_{\rm max}$ for a 
given galaxy, we will include spherical shells that are partially 
external to survey.  There will be no galaxies in this region, 
so the shell will appear to be underdense compared to its true 
density.  Thus, again, a homogeneous distribution will be 
measured to have a fractal dimension $D < 3$.  In contrast, if 
we underestimate $r_{\rm max}$, the survey is not used to 
probe large scales as well as it could be, but no bias is 
introduced.  Errors in $r_{\rm max}$, then, tend 
to result in a lower estimated fractal dimension.  This bias will 
apply on all scales if there are errors on $r_{\rm max}$ 
for all the galaxies.  
It is thus important to measure the distance to the edge of the 
survey as accurately as possible for each galaxy.  

\section{Application to the Stromlo-APM catalogue}
\label{sec:stromlo}
The Stromlo-APM redshift survey \cite{Loved92,Loved96b} consists 
of $1787$ galaxies in the southern galactic polar region, 
with magnitude $\BJ \leq  17.15$, sampled at a rate of 1 in 20 
from the APM Bright Galaxy Catalogue \cite{BGC96}and the APM 
Galaxy Survey.  Clustering 
in this catalogue has been studied using conventional statistics 
by \citeN{Loveday92} and \citeN{Tadros96}, and using the conditional 
average density by SLM.  Here we show how the 
biases explained in the previous section affect the sample, 
and how we correct for these biases.   

\subsection{Determining $r_{\rm max}$}
\label{sec:mask}
\begin{figure}
\centerline{\epsfxsize = 8.0 cm \epsfbox{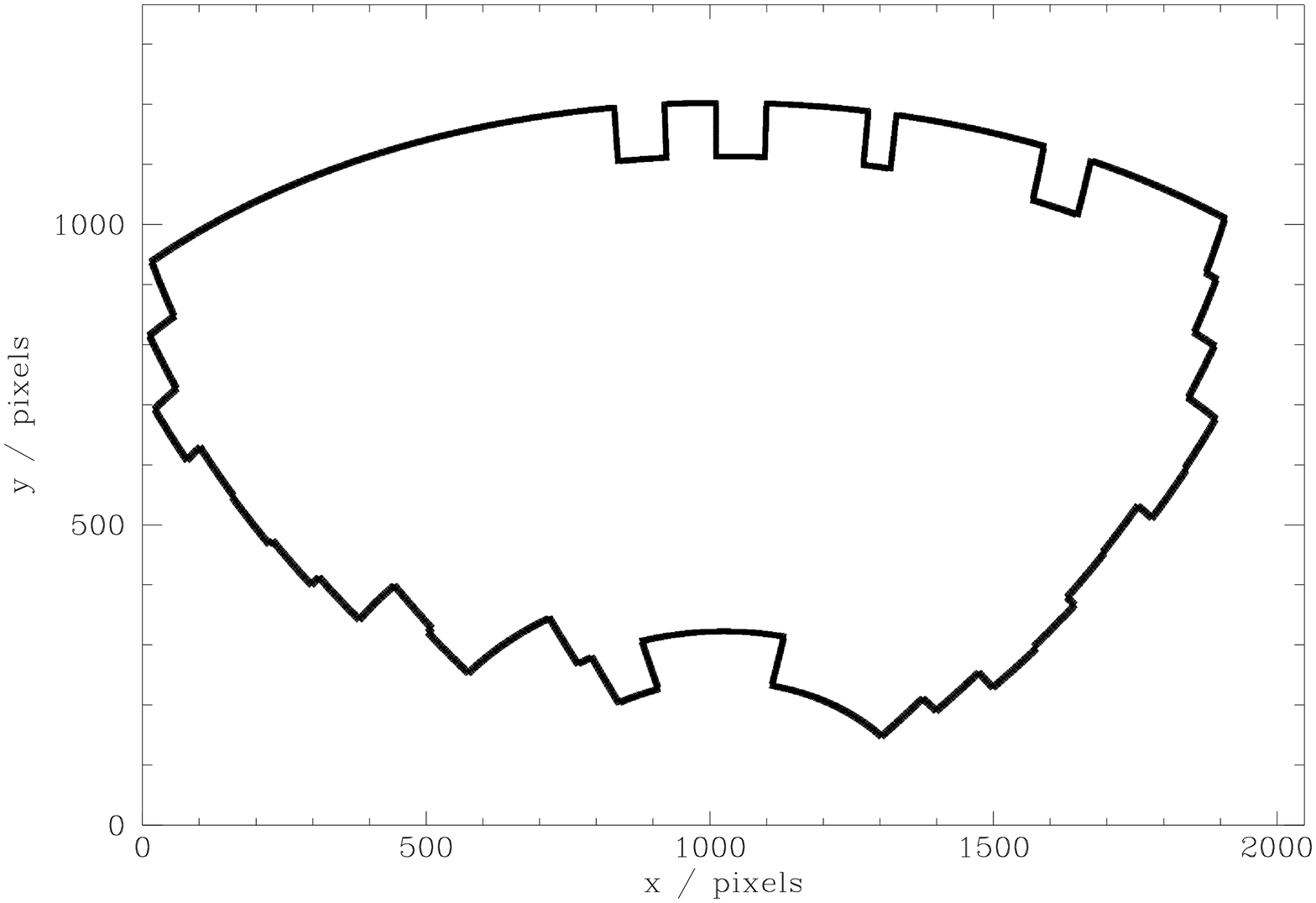}}
\caption{The boundary mask we apply to the Stromlo APM survey 
to calculate the distance of each galaxy to the edge.}
\label{fig:boundary}
\end{figure}
As explained in the previous section, it is crucial to accurately 
measure the distance from each galaxy to the edge of the survey.  
This is not a trivial task for the Stromlo-APM catalogue, since the 
survey mask is highly irregular.  We derive the angular distance 
of each galaxy to the survey perimeter by:
\begin{enumerate}
\item using the APM plate numbers to produce a pixel mask for 
the survey on a $2048\times 2048$ grid.  
\item defining pixels in this mask that are on the angular 
boundary of the survey.  
\item for each galaxy, finding the boundary pixel with the smallest 
angular separation, $\theta_{\rm min}$.
\end{enumerate}
In Fig.~\ref{fig:boundary} we show the mask that results from 
this treatment.  We find for the Stromlo data that this leads to 
a value $\theta_{\rm surv} = 22.6 \degr$ for the effective opening 
angle of the survey.  For each galaxy we restrict the pair counts 
to neighbouring galaxies within $r_{\rm max}= r\, \sin \theta_{\rm min}$, 
or $r_{\rm max}= d_{lim} - r$, whichever is the lesser.  
We only perform the final fit to the fractal dimension for points 
that are within the maximum value of $r_{\rm max}$ for each sample, 

The true survey mask in fact contains many small holes due to 
the presence of bright stars in the field.  Since galaxies may 
in fact be hidden in these holes, a conservative estimate of 
the distance to the edge of the survey may be given by the distance 
to the nearest hole.  This definition would unfortunately render the 
survey statistically useless, since it would by unable to probe 
cosmologically interesting scales.  
We suggest two possible ways of dealing with this problem, 
and both will result in a small bias.  Firstly, the holes could 
be filled with a Poisson distribution of particles, having the 
same selection function as the galaxy 
sample.  These galaxies would be homogeneously distributed, 
with a well defined average density, and so observations 
would be biased towards finding a higher fractal dimension.  
Alternatively, the holes can simply be ignored.  This will bias 
results the other way, since 
it effectively decreases the homogeneity of the distribution.  
We choose to adopt the latter scheme.  

The accurate values we find for $r_{\rm max}$ generally enable us 
to extend our model fits out to greater distances than those used 
by SLM, effectively probing the dimensionality of the Stromlo-APM 
sample at larger scales.  

\subsection{Volume limit errors}

From Fig.~\ref{fig:euclid} it will be noted that the low density 
cosmologies have redshift-distance relationships falling 
between the linear relation and that for a flat, $\Omega_0=1$ Universe.  
To estimate the 
maximum systematic error in the Stromlo-APM catalogue arising 
from the wrong assumed cosmology, we convert redshifts to distances 
using first the Euclidean, and then the $\Omega_0=1$ relationship, and 
compare the difference.  

We k-correct the Stromlo-APM data by subtracting $a(T)z$ from 
the apparent magnitude of each galaxy before calculating the 
$z_{\rm max}$ value, where $a(T)$ is a simple correction depending 
on the type, $T$, of the galaxy.  We compare this model 
for k-correction with the results we get if we do not k-correct 
at all, to get an idea of the potential systematic bias in the 
fractal dimension.  

The survey is so shallow that neither k-correction nor choice 
of cosmology have any coherent effect on estimates of the fractal 
dimension.  Although the different choices do result in some 
scatter about the mean, this is rather small compared to the 
random errors that will be presented in the next section.  

We dust-correct the sample using the extinction maps of 
\citeN{SFD98}.  This again has a negligible effect on 
measurements of the fractal dimension, since the Stromlo-APM 
survey is in a region of rather low extinction.  

We conclude, 
then, that this galaxy survey is not liable to biases incurred 
by volume limited samples being of non-constant density.

\subsection{The small-scale cut-off}
\label{sec:cutoff}
Following section~\ref{sec:poiss}, for each of the four 
galaxy samples we analyse, we create many random realizations 
of Poisson samples with the same number density and volume.  
These are analysed to find the scale at which $\Gamma^{*}(r)$ 
becomes significantly different from a flat, $D=3$ power law.  
This is used as the small-scale cut-off when fitting a 
power-law slope to the data.  

SLM use arguments involving the small-scale cut-off 
of the fractal distribution to reach the conclusion 
that one should not attempt to fit below $\< l \>$, 
the mean nearest-neighbour distance of galaxies in 
the sample.  The resulting $R_{\rm min}$ generally 
comes out rather similar to ours, as can be seen from 
a comparison with their table~1.  
 
In Table~\ref{tab:samples} we present the details of 
the four volume limited samples analyzed in this paper.  
In each case, $R_{\rm min}$ and $R_{\rm max}$ have been 
computed as described in this section.  

\begin{table}
\centering
\centerline{
\vbox {\halign {$\hfil#\hfil$&&\quad$\hfil#\hfil$\cr
\noalign{\hrule\vskip 0.03in}\cr
\noalign{\hrule\vskip 0.08in}
z_{\rm lim} & d_{\rm lim} & N       & R_{\rm min}  & R_{\rm max} \cr
\hbox{}     & \hmpc       & \hbox{} & \hmpc    & \hmpc       \cr
\noalign{\vskip 0.03in}
\noalign{\hrule\vskip 0.08in}
0.0367 & 107.2 & 320 & 2.0  & 21.8 \cr
0.059  & 169.5 & 486 & 3.2  & 33.5 \cr
0.083  & 234.5 & 402 & 10.0 & 46.6 \cr
0.097  & 271.4 & 256 & 12.0 & 54.7 \cr
\noalign{\vskip 0.08in}
\noalign{\hrule}
}}}
\caption{The limiting redshift, limiting comoving distance, 
total number of particles, and minimum and maximum scales 
probed for each of the four volume limited 
samples we consider.}
\label{tab:samples}
\end{table}

\subsection{Results}
\label{sec:res}
\begin{figure}
\centerline{\epsfxsize = 8.0 cm \epsfbox{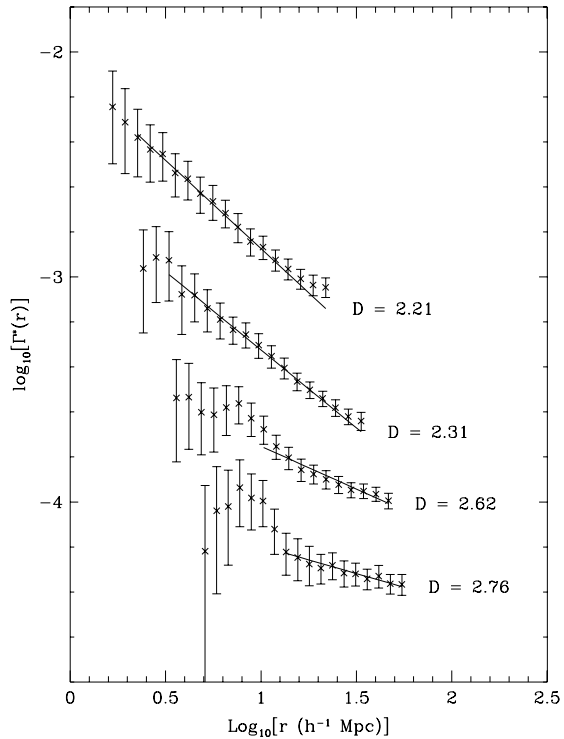}}
\caption{The measured $\Gamma^{*}(r)$ for the four samples with 
properties described in Table~\ref{tab:samples}.  Error-bars 
come from one hundred bootstrap resamples.  Also plotted are the 
best fit power-law models, shown only for the range over which 
they have been fitted.  }
\label{fig:res}
\end{figure}

By carefully restricting our method only to the 
galaxy pairs where the fractal treatment is expected 
to be valid, we are able to accurately measure the 
integrated conditional density, $\Gamma^{*}(r)$, for 
the four samples whose properties are described in 
Table~\ref{tab:samples}.  Our results are shown in Fig.~\ref{fig:res}
and Table~\ref{tab:results}.  The error bars on the data 
in Fig.~\ref{fig:res} come from bootstrap resampling of the 
galaxy sample, with one hundred bootstraps.  
\citeN{MJB92} find that, in correlation function analysis, the 
bootstrap method overestimates the error associated with each point, 
but that this overestimation is compensated for 
by the failure of a simple $\chi^2$ fit to take into 
account the covariance between bins.  We are thus justified 
in performing a $\chi^2$ fit to the data points using these 
bootstrap errors as the variances, and ignoring the 
bin-bin interdependence.  We fit a model of the form shown in 
equation~\ref{eqn:gamma_star} to the data between $R_{\rm min}$ 
and $R_{\rm max}$ 
with $A$ and $D$ as free parameters.  The error $\Delta D$ 
quoted in Table~\ref{tab:results} is the $\Delta \chi^2 = 1$ 
confidence limit, representing the 
marginalized $1$-$\sigma$ error on the parameter $D$.

As the deepest sample is found to be close to Poisson 
($D=2.76$), we expect to see the deviation from power-law 
behaviour due to small-scale bias to appear at roughly the 
same scale as was found (section~\ref{sec:poiss}) for a 
Poisson distribution with the same volume and density.  A break 
does indeed occur in the behaviour of the conditional 
density at this point, with a steeper slope on smaller scales.  
Attempting to fit a single power-law over both these regimes 
would result in significant bias in the estimated value of the 
fractal dimension.  This is also the case for the second deepest 
sample, but is less significant for the two 
shallower samples, in which the fractal dimensions are lower, 
and hence the scales at which the bias affects the conditional 
density are significantly smaller than their Poissonian values.

Note that, despite being an integrated quantity, $\Gamma^{*}(r)$ 
at times changes quite rapidly with scale: this is especially evident 
in the deeper samples.  The fractional change in volume as one 
goes from one sphere to the next is given by
\[
\frac{\Delta V}{V} \simeq \frac{3 r^2 \Delta r}{r^3} \simeq 3 \Delta \ln r = 3 \ln 10 \Delta \log r.  
\]
Since $\Delta \log r \approx 0.06$ , this implies that $\Delta V 
\approx 0.4 V$, so the volume increases by around $40$ per cent 
in each successive bin of Fig.~\ref{fig:res}.  Effectively, then, 
$\Gamma^{*}(r)$ can respond quite quickly to changes in the 
underlying slope of $\Gamma (r)$.  

\begin{table}
\centering
\centerline{
\vbox {\halign {$\hfil#\hfil$&&\quad$\hfil#\hfil$\cr
\noalign{\hrule\vskip 0.03in}\cr
\noalign{\hrule\vskip 0.08in}
z_{\rm lim} & D & \Delta D \cr
\noalign{\vskip 0.04in\hrule\vskip 0.08in}
0.0367 & 2.21 & 0.14 \cr 
0.059  & 2.31 & 0.11 \cr
0.083  & 2.62 & 0.07 \cr
0.097  & 2.76 & 0.10 \cr
\noalign{\vskip 0.08in}
\noalign{\hrule}
}}}
\caption{The fractal dimension for each of the four samples.  
The quoted uncertainties are $1$-$\sigma$ errors on $D$.}  
\label{tab:results}
\end{table}

\subsection{Comparison with mock catalogues}
In order to compare these results with the expected behaviour 
of the fractal dimension under the assumption of a conventional, 
CDM-variant cosmology, we have created mock Stromlo-APM redshift 
surveys using the approach detailed by \citeANP{CHWF} (1999, hereafter 
CHWF) for the construction of mock 2dF and SDSS redshift catalogues.  
Analysis of these simulations, using exactly the same techniques 
that were previously applied to the real data, also provides a check 
of our methods.   
The details of simulations used and biasing methods employed 
are explained in CHWF.  Briefly:
\begin{enumerate}
\item the mock catalogues are drawn from ten independent $\tau$CDM 
simulations, with $\Omega_0 = 1$, $\Lambda_0=0$.  
\item the underlying power spectrum is a \citeN{BBKS} model with 
shape parameter $\Gamma = 0.25$, and amplitude $\sigmame = 0.55$.  
\item the simulations are biased using CHWF model~1, to obtain a 
similar level of galaxy clustering to that observed in the APM 
galaxy survey,  \ie $\sigmage \approx 0.96$ \cite{MES96}.   
\item the selection function used is based on a \citeN{Schechter76} 
luminosity function with parameters taken from \citeNP{Loved92} 
($\alpha=-0.97$, $M^\star=-19.5$, $\phi_\star=1.4\times 10^{-2} 
h^3 {\rm Mpc}^{-3}$).
\item the catalogues are constructed using the Stromlo-APM mask derived in 
section~\ref{sec:mask}, and thus have the same problems of 
non-uniform geometry and holes as the true sample.
\end{enumerate}

There is one key difference between our mock catalogues and 
the real sample, namely that our assignment of luminosities 
to galaxies pays no attention to the environment of the galaxy.  
Thus, the clustering for all luminosity classes of galaxies 
will be the same in our catalogues.  This is not necessarily 
the case for the real sample.  \citeN{Loveday95} find that 
sub-$L_\star$ galaxies in the Stromlo-APM cluster more weakly 
by a factor of two than $L_\star$ galaxies, but that galaxies 
with higher luminosities than this do not show any increase 
in clustering strength.  Since $L_\star$ galaxies can be seen 
up to $\approx 180\hmpc$, our smallest sample will contain 
a significant number of fainter galaxies, and therefore may be 
expected to appear more homogeneous, relative to the mock 
catalogues, than the deeper samples.   This will not be an 
important effect for the deeper samples, which only contain 
galaxies brighter than $L_\star$.  

\begin{table}
\centering
\centerline{
\vbox {\halign {$\hfil#\hfil$&&\quad$\hfil#\hfil$\cr
\noalign{\hrule\vskip 0.03in}\cr
\noalign{\hrule\vskip 0.08in}
z_{\rm lim} & D & \Delta D \cr
\noalign{\vskip 0.04in\hrule\vskip 0.08in}
0.0367 & 2.41 & 0.12 \cr 
0.059  & 2.40 & 0.10 \cr
0.083  & 2.64 & 0.08 \cr
0.097  & 2.73 & 0.09 \cr
\noalign{\vskip 0.08in}
\noalign{\hrule}
}}}
\caption{The fractal dimension measured from mock 
Stromlo-APM catalogues.  
The quoted uncertainties the standard deviations of the results 
about the mean.}  
\label{tab:res_mock}
\end{table}
We use the same volume limits as applied to the real Stromlo-APM 
sample, and present our results in Table~\ref{tab:res_mock}. 

The use of ten mock catalogues enables us to discern firstly 
whether the bootstrap errors on the fractal dimension are 
reasonable estimates on the uncertainty of this quantity, and 
secondly whether the behaviour seen in the Stromlo-APM sample 
is consistent with the conventional clustering scenario used 
to construct the mocks.  
It will be seen from a comparison of 
Tables~\ref{tab:results} and~\ref{tab:res_mock} that both 
these points are satisfied.  The bootstrap errors do indeed 
provide a good estimate on the uncertainty in $D$, and the 
value of the fractal dimension as a function of scale shows 
excellent agreement in the real sample to that taken from 
the CDM simulations.  

We conclude that the use of the fractal 
dimension as a measure of the homogeneity of the galaxy sample 
in the Stromlo-APM catalogue in no way discriminates against 
the conventional picture of clustering.   The previously 
claimed scale-invariant fractal behaviour (SLM, $D = 2.1 \pm 0.1$)
is ruled out at the level of several-$\sigma$ for the deepest 
galaxy sample.  

\section{Analysis of larger samples: SDSS mock catalogues}
It was shown in the previous section that the Stromlo-APM 
catalogue is fairly robust to errors in dust correction, 
k-correction and assumed cosmology, since it is at low redshift 
and in an area of very low extinction.  To illustrate the 
biases that occur in a deeper sample, will now examine one 
SDSS mock catalogue, based on the CHWF $\tau$CDM simulation, 
and four variants:
\begin{enumerate}

\item {\bf MAP.  } We add extra long-wavelength power, to this 
catalogue, via the Mode Adding Procedure \cite{Tormen96,Cole97}, 
as described in CHWF.  

\item {\bf $\mathbf{\Lambda}$CDM.  } The catalogue is drawn from a 
$\Lambda$CDM simulation with 
$\Omega_0=0.3$, $\Lambda_0=0.7$.  This simulation has the same 
initial phases as the $\tau$CDM one, and so samples the same 
structures.

\item {\bf Dust.  } The galaxies have their magnitudes 
lowered by the amount corresponding to predictions from the dust maps 
of \citeN{SFD98}.  In this case, we boost the selection function 
used in creating the catalogue such that the number density is the 
same as for the original catalogue.   

\item {\bf Evolution.  } We use the strong evolution model of CHWF, 
in which the effects of evolution are not cancelled out by those 
of k-correction.  Again, the particle number is conserved by 
changing the amplitude of the selection function.  

\end{enumerate}

\begin{figure}
\centerline{\epsfxsize = 8.0 cm \epsfbox{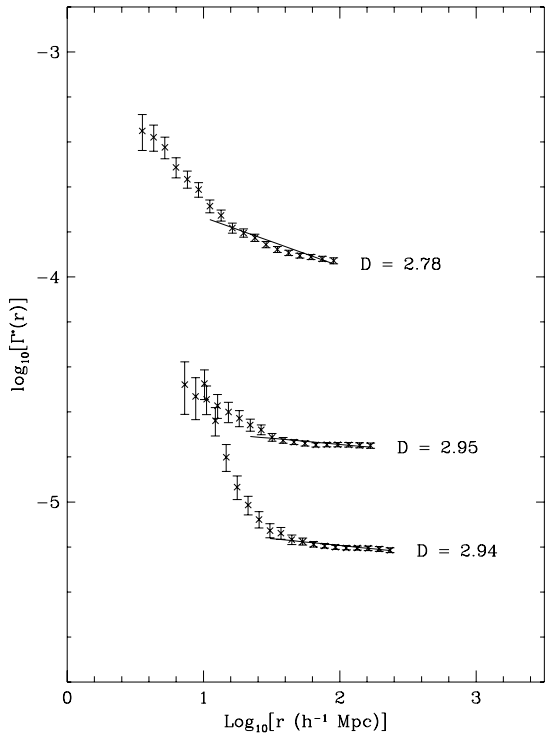}}
\caption{The measured $\Gamma^{*}(r)$ for the three 
samples from the straightforward SDSS mock catalogue.  
}
\label{fig:sdss}
\end{figure}

Apart from the dust catalogue, the construction of these variants is 
described in detail in CHWF.  Since these catalogues all effectively 
sample the same region of space, any differences between their 
properties will generally be systematic rather than random.  
We find that we are able to obtain accurate results out to 
a redshift limit of $z \approx 0.3$.  Fig.~\ref{fig:sdss} illustrates 
the behaviour of $\Gamma^{*}(r)$ for the main SDSS mock catalogue.  
The ability of this sample to probe homogeneous scales is quite evident.   
Note that we do not apply the same criterion here as used in 
section~\ref{sec:cutoff} for determining the small-scale cut-off.  
The higher sampling rate in the SDSS would result in an attempt 
to fit a power-law over a wide range of scales, but we do not 
expect this to be a good fit to the data, since the simulations 
are constructed with a CDM-like power spectrum which turns over 
to homogeneity on large scales, and they follow clustering into 
the non-linear regime on small scales.  In this instance, we are 
not concerned with finding the large-scale behaviour but obtaining 
an estimate of how well the SDSS will be able to identify homogeneity 
if it exists, and of the magnitudes of the various biases in the data.  
Hence,  we look for the large-scale, 
asymptotic behaviour of the distribution, and only fit down to 
scales where the data are still consistent with this slope.  
We present results 
in Table~\ref{tab:sdss_res} for three volume limited samples, 
$z_{\rm lim} = 0.1$, $0.2$, and $0.3$.

\begin{table}
\centering
\begin{center}
\begin{tabular}{lccc}
\noalign{\vskip -0.30in}\cr
\noalign{\hrule\vskip -0.11in}\cr
\noalign{\hrule\vskip 0.08in}
Name & \multicolumn{3}{c}{$D$}                       \cr
     & $z_{\rm lim} = 0.1$  & $0.2$ & $0.3$        \cr    
\noalign{\vskip 0.04in\hrule\vskip 0.08in}
Mock                   & 2.78 & 2.95 & 2.94  \cr
MAP                    & 2.81 & 2.96 & 2.97 \cr
$\Lambda$\hbox{CDM}    & 2.81 & 2.94 & 2.98 \cr
Dust                   & 2.82 & 2.93 & 2.95 \cr
Evol.                  & 2.81 & 2.98 & 2.91 \cr
\noalign{\vskip 0.08in}
\noalign{\hrule}
\end{tabular}

\end{center}
\caption{The observed fractal dimensions from the SDSS mock 
catalogue and four variants.  $D$ is shown for three limiting 
redshifts, $z=0.1$,$0.2$,$0.3$.  The typical error on $D$ from 
bootstrap resampling is $\pm 0.02$.   }  
\label{tab:sdss_res}
\end{table}

The first thing to note from this table is the fact that the 
sample with $z_{\rm lim}=0.1$ is consistent with the fractal 
dimension of the deepest Stromlo-APM samples analysed earlier, 
as expected since the redshifts are very similar, and the strength 
of clustering in the two catalogues is the same.  Secondly, 
the samples at $z_{\rm lim}=0.2$ are extremely close to the 
$D=3$ result expected if homogeneity has been reached.  
 
It should be noted that one of the disadvantages with catalogues 
drawn from these simulations is that they contain no power on 
scales larger than the box size, $345.6\hmpc$.  Thus, for samples 
approaching this depth, we expect to see a turnover to homogeneity 
that may be in excess of that for a CDM model with power on larger 
scales.  We analyse the MAP catalogue, which contains power on much 
larger wavelengths than this, in order to get an idea of 
the systematic difference introduced by this extra 
power.  In fact, it will be seen from comparison of the first two 
rows of Table~\ref{tab:sdss_res} that the inclusion of this power 
has negligible effect on the fractal dimension.   

Judging from Fig.~\ref{fig:euclid}, we might expect cosmology to be a 
major factor in biasing deeper the samples from the $\Lambda$CDM 
catalogue since the redshift-distance relations are quite discrepant 
at these redshifts.  This turns out not to be the case.  The assumption 
we have made, of an $\Omega_0 = 1$ cosmology, should result 
in an artificial squashing of the volume element at high redshift.  
Thus, the galaxy density would be expected to increase with 
redshift in a volume limited sample.  However, the 
change in limiting redshift of each particle, described by 
equation~\ref{eqn:dmod} has the opposite effect: the value 
of $d_{\rm max}$, the maximum distance (in comoving co-ordinates) 
at which a particular galaxy can be seen by the magnitude limited 
survey, 
is reduced under the $\Omega_0 = 1$ assumption, with the effect 
of reducing the density at high redshift.  
The net effect of these two considerations is thus much smaller 
than either of them taken singly, and the catalogue analysed with 
the wrong cosmology does not produce significantly different 
results from the main catalogue.  Note that this means we have, 
on average, intrinsically brighter galaxies at high redshift in 
the sample.  This is not a problem in our mock catalogues, since 
there is no dependence of clustering properties on galaxy luminosity, 
but could result in a bias for a real survey.  

Similarly, as shown in the final two rows of Table~\ref{tab:sdss_res}, 
introducing dust and using the wrong model for the effects of 
k-correction and evolution appear to have little 
or no appreciable effect on the fractal dimension of the sample 
at any redshift.  This is despite the fact that the SDSS survey 
extends to quite low galactic latitudes at its southern-most 
extremities.  

We conclude that for the SDSS sample, despite its much greater 
depth and angular coverage than the Stromlo-APM, there is 
unlikely to be serious 
systematic bias caused by any of the volume limit effects outlined in 
section~\ref{sec:vl_fx}.

\section{Discussion}
\label{sec:conc}
We have shown that, contrary to previous results, 
the distribution of galaxies in the Stromlo-APM redshift survey 
approaches homogeneity as the sample depth is increased.   
Galaxies in volume-limited subsamples from the Stromlo-APM catalogue 
generally cluster as fractal distributions, with higher fractal 
dimension on for deeper samples.  

At the deepest scale that can be reliably probed ($\approx 60\hmpc$),  
the conditional density of behaves as a power-law, with slope 
given by a fractal dimension $D=2.76 \pm 0.10$, close to the 
value of $3$ expected if the Universe is homogeneous.   
Whilst this is not proof of complete homogeneity on larger scales, 
we note that:
\begin{enumerate}
\item the value for the fractal dimension is consistent with that 
expected from a conventional CDM model of galaxy clustering, using 
parameters for the shape and amplitude of the power spectrum that 
have been measured from the Stromlo-APM sample itself.  
\item this value is inconsistent, at the several-$\sigma$ level, 
with previous results finding $D=2.1 \pm 0.1$ (SLM).  
\end{enumerate}

We have shown that the fractal dimension measured from the 
Stromlo-APM survey and indeed future, deeper surveys like 
the SDSS is generally unaffected by reasonable errors 
in k-correction, dust correction, and assumed cosmology.  
Why, then, do our results differ from previously published 
work?  The accuracy of our method for estimating the distance 
of a galaxy to the edge of the survey, as presented in 
section~\ref{sec:mask}, results in a two-fold gain in 
probing large-scale inhomogeneities.  Firstly, we are confident 
that there are no errors biasing the fractal dimension on 
large scales, and, secondly, this confidence enables us to 
measure the conditional density out to scales around thirty 
per cent deeper than SLM, where the distribution is closer to 
homogeneity.

\subsection{Application to PSCz}
A far more immediate prospect than the SDSS is the application of these 
techniques to the PSCz survey.  This dataset has the advantage that it 
has a large angular coverage, so, despite having similar depth to 
the Stromlo-APM, its usefulness in measuring the conditional density 
at large scales rivals the SDSS.  An important factor in extracting the 
best information from the PSCz will be dealing with its irregular 
selection geometry.  The power of 
a fully spherically-symmetric survey to probe the largest scales in 
a statistically valid way is enormous, but the PSCz is restricted by 
lack of data at low galactic latitudes.  Selecting 
the largest sphere that can be placed in one hemisphere without 
intersecting with this zone of avoidance reduces the largest scale 
probed by approximately a factor of two, but the resulting 
$r_{\rm max}$ is still large enough that we should expect to see 
a $D=3$ dimensionality if the CDM scenario is valid.  
Further complications arise due to the missing strip of IRAS data 
that results in a hole in the PSCz data.  Since this strip is 
basically orthogonal to the zone of avoidance, including it 
in restricting the size of the sphere results in a severe 
additional reduction in the maximum scale that can be probed.  
We suggest that, in order to maximise the useful range of the 
data, each hemisphere should by analyzed with the strip unfilled, 
resulting in a bias away from homogeneity, and then padded with 
a Poisson sample of particles, resulting in a bias towards homogeneity.  
These extremes should successfully bracket the true behaviour of 
the sample, resulting in a quantified systematic error on the 
derived fractal dimension.

\section*{Acknowledgements}
The author acknowledges the support of a PPARC studentship and funding 
from Durham University.  The author would like to thank Shaun Cole 
and Carlton Baugh for useful suggestions during the final stages of 
this work.  Thanks also to the referee, Luigi Guzzo, for a prompt 
and detailed report.  

\bibliographystyle{mnras}

\end{document}